# Back-of-the-Envelope Computation of Throughput Distributions in CSMA Wireless Networks

Soung Chang Liew, Caihong Kai
Dept. of Information Engineering
The Chinese University of Hong Kong
Hong Kong SAR, P.R.China
soung@ie.cuhk.edu.hk;chkai6@ie.cuhk.edu.hk

Jason Leung, Bill Wong
Altai Technologies
Unit 105, 1/F, Block 9, Hong Kong Science Park
Hong Kong SAR, P.R. China
hcleung@altaitechnologies.com; billwong@altaitechnologies.com

*Abstract*

**This work started out with our accidental discovery of a pattern of throughput distributions among links in IEEE 802.11 networks from experimental results. This pattern gives rise to an easy computation method, which we term back-of-the-envelop (BoE) computation, because for many network configurations, very accurate results can be obtained within minutes, if not seconds, by simple hand computation. BoE beats prior methods in terms of both speed and accuracy. While the computation procedure of BoE is simple, explaining why it works is by no means trivial. Indeed the majority of our investigative efforts have been devoted to the construction of a theory to explain BoE. This paper models an ideal CSMA network as a set of interacting on-off telegraph processes. In developing the theory, we discovered a number of analytical techniques and observations that have eluded prior research, such as that the carrier-sensing interactions among links in an ideal CSMA network result in a system state evolution that is time-reversible; and that the probability distribution of the system state is insensitive to the distributions of the "on" and "off" durations given their means, and is a Markov random field. We believe these theoretical frameworks are useful not just for explaining BoE, but could also be a foundation for a fundamental understanding of how links in CSMA networks interact. Last but not least, because of their basic nature, we surmise that some of the techniques and results developed in this paper may be applicable to not just CSMA networks, but also to other physical and engineering systems consisting of entities interacting with each other in *time* and *space*.**

*Keywords*: **CSMA; 802.11; WiFi; multiple access; on-off process; telegraph process.**

## 1. INTRODUCTION

This paper concerns the computation of throughput distributions of links in carrier-sense multiple-access (CSMA) wireless networks, such as the IEEE 802.11 networks. While methods for throughput computation of CSMA networks now appear in standard textbooks, most, if not all, textbook materials deal with the case in which all links can sense each other in an all inclusive manner. More recently, Bianchi [1] provided the analysis for 802.11 networks, again assuming "all-inclusive carrier sensing".

With the widespread deployment of 802.11 networks, it is now common to find numerous co-located 802.11 networks. The carrier-sensing relationships among the links of these networks are **non-all-inclusive** in that not all links can sense each other. It is extremely difficult to extend the method in [1] to the non-all-inclusive case. Appendix A details the issues involved.

Observing this, Ng and Liew [2, 3] provided an approximate method for non-all-inclusive CSMA networks. The method makes use of a set-theoretic inclusion-exclusion principle to approximate the overlapped airtime usage between adjacent links. Yan *et al.* [4] later refined the method by incorporating the consideration of packet collisions. The methods in [2, 3, 4] are analytical but not algorithmic in nature. Appendix A shows that the results obtained by these methods are not as accurate as the simpler method presented in this paper.

There have been numerous publications on non-all-inclusive carrier-sense networks and this is indeed a "hot topic" among researchers. Besides [2, 3, 4], other recent work includes [5, 6, 7], from which earlier work can be traced. The method in this paper is simpler than the prior methods. In addition, in this paper, we attempt to explain our method with a precise mathematical model: specifically, an **ideal CSMA network** is modeled as a set of **interacting on-off telegraph processes**. By adding rigor, we hope to set up a framework for future extension work, as well as to unearth unknown fundamental properties of CSMA networks.

We refer to the method in this paper as a back-of-the-envelop (BoE) method because for networks of modest size, the results can be obtained in a matter of minutes, if not seconds, with simple hand computation. Take the network in Fig. 1. The carrier-sensing relationship is described by the contention graph shown, in which links are represented by vertices, and an edge joins two vertices if the transmitters of the two associated links can sense each other. The "normalized" throughput distributions of the links, $(Th_1\ Th_2\ Th_3\ Th_4)$, can be quickly approximated to be $(1\ 0\ 0.5\ 0.5)$ within seconds.

The next section will specify the BoE method formally. To illustrate its simplicity, for the time being we describe its mechanic without justification – even an elementary school kid could follow this mechanic for networks of small size. For Fig. 1, the BoE computation proceeds as follows. With respect to the contention graph, we try to put a label of 1 to as many vertices as possible with the constraint that two vertices joined by an edge cannot both be 1 (1 represents transmission, and two links that can sense each other cannot transmit together). The "greedy states" are (1 0 1 0), (1 0 0 1), and (0 1 0 0). The states (1 0 1 0) and (1 0 0 1) have two 1's while the state (0 1 0 0) has only one 1. We retain only the states with the maximum number of 1's, which are (1 0 1 0) and (1 0 0 1) for the network. We then add the vectors together and then divide the sum by the number of vectors, yielding (1 0 0.5 0.5). And these are the normalized throughputs $(Th_1\ Th_2\ Th_3\ Th_4)$.

Historically, we were led to the BoE computation method by simulation and real-network experimental results. The theory of interacting on-off processes presented in this paper was developed thereafter to explain the experimental observation. In developing the theory, we discovered a number of analytical techniques and observations that have eluded prior research, such as that the system state of an ideal CSMA network evolves in a **non-Markovian** manner **in time** but is **time-reversible**; that the system-state probability distribution is **insensitive** to the distributions of the countdown and transmission times given the ratio of their means; and that the system-state probability distribution is a **Markov random field** [8].

From a "resource allocation" standpoint, the CSMA protocol is an implementation of a specific **distributed resource allocation** discipline, with an implicit utility objective. BoE is just computing the resulting throughput distributions as dictated by the implicit utility objective. It is therefore not surprising that the theoretical framework constructed by this paper to explain BoE also provides a foundation to understand the distributed resource allocation as implemented by the CSMA protocol. More comments on resource allocation will be presented in section 5.

The theory yields much insight into the inner working of CSMA networks. For example, an insight is that in an 802.11 network with $L$ links competing in a greedy manner, there is a maximum of $2^L$ possible states in terms of who is transmitting and who is not; but only very few of these states are probable, and the probable states are equally probable. Our results also indicate that link starvation (e.g., link 2 in Fig. 1) is prevalent in CSMA networks. A practical application of BoE is for quick identification of problems in a network so that remedies could be devised.

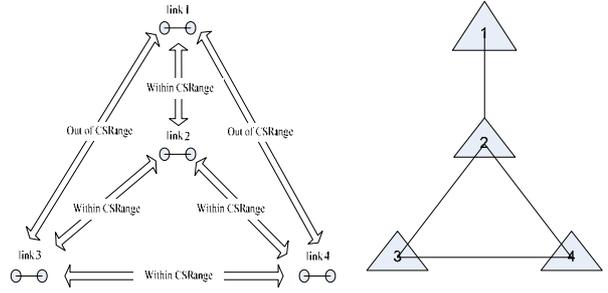

Fig.1. An example network and its associated contention graph.

The roadmap of our presentation and a **synopsis of the logical relationships of our results** are as follows. If desired, the reader could skip the next four paragraphs and come back later to get an overall big picture as the individual sections are read.

In section 2, we state the method of BoE formally. We provide simulation and real-network experimental results to demonstrate its accuracy. In section 3, we show that BoE can be explained by treating the on-off process of a link as a Markov process in which the countdown and transmission times are exponentially distributed, and then letting the ratio of mean countdown time to mean transmission time go to zero. Our experiments, however, did not assume exponential distributions of countdown and transmission times, and yet the BoE results are highly accurate. This gave us a strong hint that the results are insensitive to the form of the distributions, and only the ratio of their means matter.

Section 4 is devoted to proving the **insensitivity** property. Several other key properties of **interacting on-off processes** are also proven along the way, including the fact that the system state defined by the set of "on" processes is **time-reversible** and that its stationary state distribution is a **Markov random field**. The proof of insensitivity in section 4 requires us to make an **invariant residual-time distribution assumption**. The assumption states that that the residual countdown time and residual transmission time of a link at the state-transition epochs of other links has the same distributions as those observed at a random point in time. Our simulation experiments did not make this assumption, and yet the insensitivity results hold in all the experiments we conducted. This suggests that the invariant residual-time distributions are an intrinsic property of the system. In other words, they are "corollaries" rather than an "assumptions".

Following this hint, appendix B approaches the whole problem from a different angle using a **continuous state-space** treatment. Appendix B not only proves the invariant residual-time property, but is also a self-contained proof for the insensitivity result. That is, one could in principle go directly to the pure mathematical treatment of appendix B without the steps in section 4. However, the approach in section 4 arguably gives us more physical insights as to what happens in the system. Appendix C provides yet another approach to the problem using the **mixture-of-gamma-distributions** method.

The main focus of this paper is the BoE method and the theoretical underpinning of it. Section 5 is a brief discussion of the implications and applications of the results. Thanks to its simplicity, we believe that BoE and its variants could find use in engineering designs in many ways beyond those discussed in this paper. In particular, BoE allows us to make shortcuts in our evaluation of the system performance.

## 2. BASIC BoE COMPUTATION AND EXPERIMENTAL CONFIRMATION

The following formalizes the description of BoE:

**BoE Computation**
1. Draw the contention graph.
2. Identify the maximal independent sets.
3. Retain only the maximum independent sets (MIS).
4. The normalized throughput of link $i$ is $n_i/n$, where $n$ is the number of MIS identified in step 3, and $n_i$ is the number of MIS in which link $i$ appears.
5. Convert normalized throughputs to throughputs in bps.

An independent set is a subset of vertices such that no edge joins any two of them; a maximal independent set is an independent set that is not the subset of another independent set; a maximum independent set (MIS) is an independent set with the maximum cardinality [9]. We note that counting MIS is an NP-complete problem, and therefore BoE can get out of hand for networks of large size. For networks of small size, such as 802.11 networks within a building, the problem is manageable. We briefly speculate on whether we can get away from the NP-complete problem in section 5.

Refer to Fig. 1 again. After step 2, the maximal independent sets identified are (1 0 0 1), (1 0 1 0), (0 1 0 0) in state notation. After step 3, only the MIS (1 0 0 1), (1 0 1 0) remain. Step 4 determines the normalized throughput distributions to be (1 0 0.5 0.5). In step 5, a normalized throughput of 1 corresponds to the throughput of a link transmitting in isolation of other links, as if it were the only link in the whole network. For example, for 802.11b, after taking into account the various header, countdown, and ACK overheads, the throughput of an isolated link for a UDP session is

$Th_{\text{single link}} = 6.06\text{Mbps}$; 30.91Mbps (for 802.11b; 802.11a)

The actual throughput of a link with a normalized throughput of $Th_{\text{norm}}$ is then computed as

$Th_{\text{actual}} = Th_{\text{norm}} \cdot Th_{\text{single link}}$

Fig. 2 shows the results of BoE computation for a number of network topologies, as well as the corresponding NS2 simulations results for UDP and TCP sessions. Typical 802.11b parameters were used in the simulations: (i) data rate and basic rate of 11Mbps and 1 Mbps, respectively; (ii) packet payload of 1460 Bytes; (iii) for UDP session, CBR flow of 7 Mbps to saturate the network. $Th_{\text{single link}}$ for a TCP session is 4.84Mbps. As can be seen, the accuracy of BoE is quite amazing for such a simple method. Simulations of many topologies other than those in Fig. 2 also bear out BoE. As mentioned earlier, we were originally led to BoE from observations of simulation and real-network experimental results rather than from analytical deduction. The structural simplicity of BoE led us to believe that there might be a deeper underlying theory. Sections 3 and 4 detail our attempts to unveil the secret behind BoE.

Besides simulations, real network experiments also bear out BoE. We set up two topologies ((1) and (2) in Fig. 2) with four pairs of DELL Latitude D505 laptops PCs with 1.5GHz Celeron Mobile CPU. Each node has a NETGEAR WAG511GE Dual Band Wireless PC card, and run Fedora5 with MADWifi driver. All Atheros chipset extensions are disabled. The experiments were conducted outdoor on 802.11a channel 36. The throughput of an isolated link is around 29Mbps. As shown in Fig. 2, the experiment results match well with BoE's prediction. In the real environment, we found it difficult to totally isolate two links to keep them out of carrier-sensing range. This is the reason why the measured throughput distributions are not as extreme as predicted.

| (1) 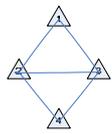 | BoE | (1, 0, 0, 1) | (2) 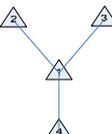 | (0, 1, 1, 1) |
|---|---|---|---|---|
| | NS2, UDP | (0.96, 0.02, 0.02, 0.97) | | (0.03, 0.99, 0.98, 1) |
| | NS2, TCP | (0.97, 0.01, 0.01, 0.97) | | (0.02, 0.99, 1, 1) |
| | Real network | (0.76, 0.10, 0.10, 0.74) | | (0.1, 0.8, 0.8, 0.8) |
| (3) 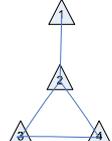 | BoE | (1, 0, 0.5, 0.5) | (4) 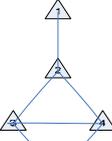 | (0.75, 0.25, 0.25, 0.25, 0.5) |
| | NS2, UDP | (0.99, 0, 0.50, 0.51) | | (0.75, 0.26, 0.26, 0.26, 0.51) |
| | NS2, TCP | (0.99, 0, 0.49, 0.51) | | (0.74, 0.25, 0.24, 0.25, 0.50) |
| (5) 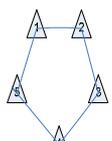 | BoE | (0.4, 0.4, 0.4, 0.4, 0.4) | (6) 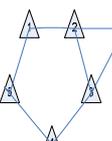 | (1, 0, 0, 1, 0, 1) |
| | NS2, UDP | (0.41, 0.40, 0.41, 0.40, 0.40) | | (0.92, 0.04, 0.04, 0.92, 0.04, 0.93) |
| | NS2, TCP | (0.40, 0.39, 0.40, 0.40, 0.40) | | (0.95, 0.02, 0.02, 0.95, 0.03, 0.95) |
| (7) | BoE | (0.5, 0.5, 0.5, 0.5, 0.5, 0.5) | (8) | (0.2, 0.4, 0.4, 0.8, 0.6, 0.6) |
| (graph) | NS2, UDP | (0.5, 0.5, 0.49, 0.5, 0.5, 0.5) | (graph) | (0.20, 0.41, 0.41, 0.80, 0.60, 0.60) |
| (graph) | NS2, TCP | (0.46, 0.51, 0.51, 0.46, 0.46, 0.51) | | (0.21, 0.39, 0.4, 0.79, 0.61, 0.60) |

Fig. 2 Contention graphs of various network topologies and the BoE computed and experimental results for them.

## 3. EXPLAINING BoE

It turns out that steps 1-4 of BoE for normalized throughput computation implicitly assumes an idealization of the CSMA system, and step 5 for actual throughput computation can be considered as a perturbation step that builds on top of the idealized outcome. We focus on steps 1-4 here. The last paragraph of section 3 briefly discusses the perturbation analysis. Not long after BoE was discovered from experimental observations, we quickly realized two essential underpinnings of steps 1-4, as embodied in the following propositions.

**Proposition 1:** The system spends most of its time in MIS, and very little time in other states.

**Proposition 2:** The MIS states are equally likely. That is, the system spends approximately equal amounts of time in each MIS.

Step 3 of BoE makes the rough approximation that the system spends zero time in non-MIS. Step 4 of BoE implicitly makes use of Proposition 2. Much of sections 3 and 4 explain the idealized conditions leading to these two propositions, hence BoE. In doing so, we also discover a number of other more general results.

### 3.1 Quick Review of 802.11 CSMA Protocol

Although our discovery of BoE is from 802.11 network experiments, BoE is a technique that can be used for any CSMA network. There are three basic features to the generic CSMA protocol: (1) carrier sensing to avoid collisions; (2) random countdown to minimize collisions; (3) backoff upon a collision to regulate transmission attempt rate. This section briefly reviews 802.11 CSMA [10] as a specific instance of how these features are implemented, as well as to lay down the context for later presentation of an **ideal** CSMA network model.

In 802.11, a station that has packets to send must first sense the channel to be idle for a duration of DIFS (Distributed Interframe Spacing) plus an integral number of backoff timeslots before transmitting a packet. For each transmission attempt, the station chooses a random integer back-off value uniformly distributed in the range of [0, CW], where CW is the so-called contention window. For a new packet with no prior collisions, CW is initially set to $CW_{min}$. The back-off value is decremented by one for each slot the channel is sensed idle. If the channel is sensed busy before the counter reaches zero, the decrementing process is frozen until the channel is sensed idle for a DIFS period. After transmitting a packet, the sender expects to receive an acknowledgement (ACK) within a SIFS (Short Inter Frame Spacing) period. If an ACK is not received within ACK, timeout occurs and the packet is assumed to be lost, and CW is doubled for the retransmission attempt. Upon successful transmission, the CW is reset to $CW_{min}$ for the next packet.

For an isolated link, the time consumed by a successful packet transmission consists of (i) DIFS; (ii) the random number of backoff timeslots; (iii) packet consisting of physical-layer preamble/header, MAC Header, and data payload; (iv) SIFS (v) ACK. For each packet, the minimum "unshared" airtime within its carrier-sensing range that must be exclusively dedicated to it is

$T_{tr}$ = PACKET + SIFS + ACK + DIFS      (1)

In addition, it also consumes a random countdown time of $T_{cd}$. Links that can sense each other cannot share $T_{tr}$. However, $T_{cd}$ can be shared because they can count down together when the surrounding medium is idle. In this paper, we define the countdown overhead as

$c = E[T_{cd}]/E[T_{tr}]$      (2)

When collisions are rare, $E[T_{cd}] \approx T_{slot} * CW_{min}/2$.

The average airtime consumed by an isolated link for each packet is $(1+c)E[T_{tr}]$. For a network with two saturated links that can sense each other and that compete on an equal basis, the average airtime consumed by two packets, one from each link, is $(2+c)E[T_{tr}]$. Note that the countdown time is not $2c$ because the two links share the same countdown time. For a general *non-all-inclusive carrier-sense* network, the situation becomes more complicated because links out of the carrier-sensing range of each other can transmit together (i.e., share airtime $T_{tr}$).

### 3.2 Ideal CSMA Link Model and Validity Conditions for BoE

In this subsection, we construct the conditions that lead to the validity of Propositions 1 and 2. To isolate the carrier-sensing effect, it behooves us to eliminate the collision effect so that we have a simpler **ideal CSMA link model** in which collisions do not occur. In a network, collision probability will increase with the number of links. If we ignore the effect of collisions, the computed throughput will be on the optimistic side. On the other hand, multiple links can also count down together even if they are within each other's carrier sensing range, and this sharing of countdown time may lead to increased throughput. These two considerations have opposing effects that may offset each other to a certain extent. For example, with all-inclusive carrier-sensing, the NS2 simulated per-link throughputs for 802.11b networks with $L$ links are $Th$ = 6.06Mbps for $L = 1$;

$Th$ = 3.25Mbps for $L = 2$; $Th$ = 2.28Mbps for $L = 3$; $Th$ = 1.77 Mbps for $L = 4$. We see that the effect of collisions has not pulled down the system throughput $Th \times L$.

One way to "model away" collisions while retaining the essence of carrier sensing is to assume a "continuous" countdown time, so that the probability of two links having the same countdown time, hence their collision probability, is zero. Thus, instead of the random *integer* number of timeslots selected from the interval [0, CW] in 802.11, our ideal CSMA link uses a random *real* number drawn from the same interval.[1] Condition 1 below embodies this modeling technique.

**Condition 1:** Countdown collisions are negligible so that countdown time can be modeled as a continuous random variable analytically.

**Condition 2:** Countdown time is negligible compared with transmission time.

**Condition 3:** The countdown time and transmission time are exponentially-distributed random variables.

Condition 3 is included to simplify things at this preliminary stage, so that we could work within the familiar construct of the Continuous-Time Markov Chain (CTMC). It will be removed in section 4 – its removal is significant because the exponential assumptions are not valid in many practical CSMA protocols (e.g., 802.11). We are now ready to explore the implications of Conditions 1, 2, and 3. We will proceed in three steps: (i) define an **ideal CSMA link** based on Condition 1; (ii) add Condition 3 to analyze the throughput distributions of the **ideal CSMA network** made up of the ideal CSMA links; (iii) add Condition 2 to arrive at the normalized throughputs as computed by BoE in its step 4.

*(i) Ideal CSMA Link/Network Model based on Condition 1*

With continuous countdown and transmission times, the stochastic process of an *isolated link* is simply the on-off telegraph process, with 1 representing transmission, and 0 representing countdown. In the ideal CSMA network, we have a set of **interacting telegraph processes**. Fig. 3 depicts a model that captures the state of a link *i*, and how it can affect and be affected by other links. In Fig. 3, $RC_i, RT_i \in [0, +\infty)$ are the remaining countdown and transmission times; and $R_i, S_i \in \{0,1\}$ are the carrier sensing input and output. The link can be in one of three possible states: 1) transmission state: when $(R_i, RC_i, RT_i) = (0, 0, +ve)$; 2) active countdown state: when $(R_i, RC_i, RT_i) = (0, +ve, 0)$; 3) frozen countdown state: when $(R_i, RC_i, RT_i) = (1, +ve, 0)$ (i.e., another link within its carrier sensing range is transmitting, and countdown is frozen).

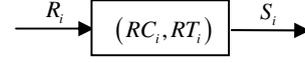

Fig. 3. The ideal link model

The state transition diagram is shown in Fig. 4. Let $T_{cd}$ be the random countdown time generated according to the probability density $p_{T_{cd}}(t_{cd}) = f(t_{cd})$, and $T_{tr}$ be the transmission time with probability density $p_{T_{tr}}(t_{tr}) = g(t_{tr})$.[2] When $RT_i$ reaches 0 from a non-zero value, a new $RC_i$ is generated; when $RC_i$ reaches 0 from a non-zero value, a new $RT_i$ is generated. At any time, either $RT_i$ or $RC_i$ must be 0, since the link is either transmitting, in which case $RC_i = 0$; or counting down, in which case $RT_i = 0$ whether the countdown is active or frozen. While in transmission state $dRT_i / dt = -1$; while in active countdown state $dRC_i / dt = -1$; $dRT_i / dt = dRC_i / dt = 0$ otherwise.

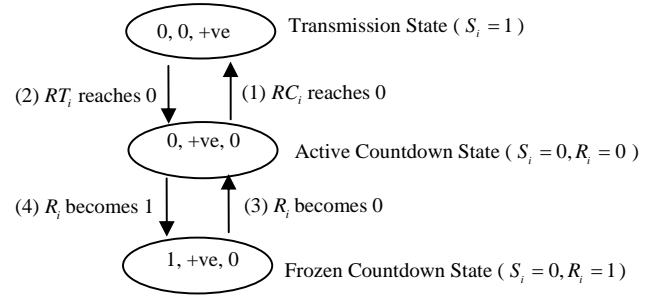

Fig. 4. State-transition diagram for an ideal link (The labels (1) to (4) of the transition-triggering events are mapped to the pseudo-code segments in subsection 4.2.1).

We could write a computer simulation module for the link model, and then connect the modules together to form the simulation program for an overall network. Such a simulation module is actually a *precise* way of specifying the **operating logic** of the ideal link. Fig. 5 shows how the link modules can be connected together for the network of Fig. 1. Subsection 4.4 will provide details of an event-driven simulation link module.

---

[1] We are not advocating reducing the timeslot size in implementation, which may not be practically viable. The ideal link model is an analytical model. It is not meant to be an implementation blueprint.

[2] Throughout this paper, we assume all links have the same countdown and transmission time distributions. Our methods can easily be extended to the general case where different links have different distributions.

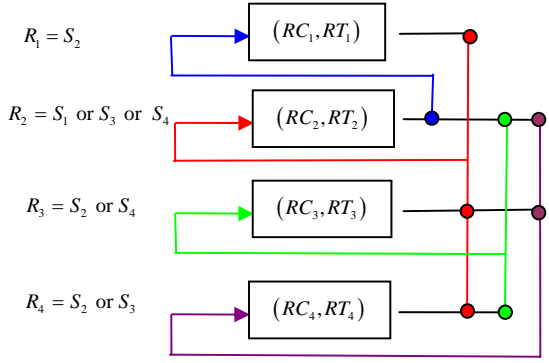

Fig. 5. The ideal network model based on interconnection of ideal links for the network in Fig. 1.

We define the **abbreviated state** of link $i$ to be $S_i$, where $S_i = 1$ if and only if $RT_i > 0$ (i.e., when the link is in the midst of a transmission). Consider a network of $L$ links. Define the **system state** as $S = S_1 S_2 \ldots S_L$. Note that given the global $S$, we know whether link $i$ is in transmission state ($S_i = 1$), active countdown state ($S_i = 0$, and $S_j = 0$ for all neighbors $j$ of $i$), or frozen countdown state ($S_i = 0$, and $S_j = 1$ for at least one neighbor $j$ of $i$). Henceforth, the term "state" simply refers to $S_i$ or $S$, unless otherwise stated.

Fig. 6 shows the state-transition diagram of the network of Fig. 5. To avoid clutters, in Fig. 6 we have merged the two directional transitions between two states into one line. Each transition from left to right corresponds to the beginning of a transmission on one particular link, while the reverse transition corresponds to the ending of a transmission on the same link. For example, the transition from 1000 to 1010 is due to the beginning of a transmission on link 3 while link 1 is transmitting; while the reverse transition from 1010 to 1000 is due to the ending of a transmission on link 3 while the transmission on link 1 continues. Since we assume Condition 1, so that $T_{cd}$ is a continuous random variable, with probability zero two links will begin or terminate transmission simultaneously (picture the on-off processes).

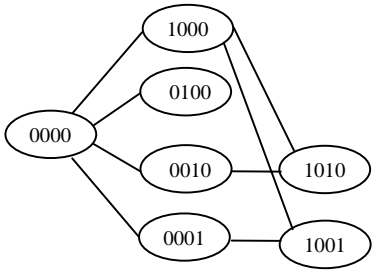

Fig. 6. The state-transition diagram for the ideal network in Fig. 5.

**Definition of State Connectivity**: Two feasible state realizations $S = s$ and $S = s'$ are said to be connected if it is possible to have a direct transition from $s$ to $s'$, and vice versa, without traversing other states.

For example, in Fig. 6, 1000 and 1010 are connected; but 1010 and 1001 are not connected.

*Observation 1*: Two feasible states $s$ to $s'$ are connected if and only if all the links transmitting in $s$ are also transmitting in $s'$, and there is one extra link transmitting in $s'$ that is not transmitting in $s$; or vice versa (i.e., their Hamming distance is 1).

**Definition of Left and Right States:** For two connected states, we refer to the state with one fewer (more) link transmitting as the left (right) state.

*(ii) Throughput Distributions based on Conditions 1 and 3*

Let $P_s = P_{s_1 s_2 \ldots s_L}$ be the fraction of time the network is in state $s = s_1 s_2 \ldots s_L$. Then the fraction of time link $i$ is transmitting is $x_i = \sum_{s: s_i = 1} P_{s_1 s_2 \ldots s_L}$, which corresponds to the normalized throughput of link $i$.

Let us now additionally assume Condition 3 that both countdown and transmission times are exponentially distributed. An outcome is that the evolution of the system becomes a Markov process. Let $1/\lambda = E[T_{cd}]$ and $1/\mu = E[T_{tr}]$. Then for any pair of connected states, the transition from the left state to the right state occurs at rate $\lambda = 1/E[T_{cd}]$, and the transition from the right state to the left state occurs at rate $\mu = 1/E[T_{tr}]$. It is easy to verify that the resulting CTMC is time-reversible and therefore detailed balance applies [11]. Specifically, for two connected states $s$ and $s'$, with $s$ being the left state and $s'$ being the right state,

$$P_s = c P_{s'} \tag{3}$$

where $c = E[T_{cd}]/E[T_{tr}] = \mu/\lambda$ is the *transition-rate ratio*, which is also the countdown overhead defined in (2).

An immediate corollary of observation 1 and (3) is that all feasible states with the same number of transmitting links (i.e, states in the same column of the state-transition diagram) have the same probability. Specifically, let $S^{(n)}$ be the subset of feasible states with $n$ transmitting links. Then,

$$P_s = B/c^n \quad \forall s \in S^{(n)}, \text{ where } B = \left( \sum_{n=0}^{L} |S^{(n)}|/c^n \right)^{-1} \tag{4}$$

As an example, applying the above to the network of Fig. 5 gives

$$P_{0000} = B = \left(1 + 4/c + 2/c^2\right)^{-1}$$
$$P_{1000} = P_{0100} = P_{0010} = P_{0001} = \left(c + 4 + 2/c\right)^{-1} \quad (5)$$
$$P_{1010} = P_{1001} = \left(c^2 + 4c + 2\right)^{-1}$$

The normalized throughputs of the links are then given by

$$x_1 = P_{1000} + P_{1010} + P_{1001} = \left(c + 4 + 2/c\right)^{-1} + 2\left(c^2 + 4c + 2\right)^{-1}$$
$$x_2 = P_{0100} = \left(c + 4 + 2/c\right)^{-1}$$
$$x_3 = P_{0010} + P_{1010} = \left(c + 4 + 2/c\right)^{-1} + \left(c^2 + 4c + 2\right)^{-1}$$
$$x_4 = P_{0001} + P_{1001} = \left(c + 4 + 2/c\right)^{-1} + \left(c^2 + 4c + 2\right)^{-1}$$
$$(6)$$

Before leaving this part, we note that (4) can in principle be obtained using the technique propounded in [12], although [12] deals with the multihop case. In [12], for analytical tractability, the dependencies of successive hops of packets are decoupled by the assumption that the transmission attempts at each node are a Poisson point process. This maps to our exponential countdown time here. This assumption, however, is not compatible with practical CSMA protocols (e.g., 802.11). In most practical CSMA protocols, the countdown process has memory and that the countdown continues where it left off after coming out of the frozen state. In section 4, we show that (4) remains valid even if that is the case.

*(iii) Throughput Distributions based on Conditions 1, 2, and 3*

For situations where $E[T_{cd}] \ll E[T_{tr}]$, or when we simply want to make a quick approximation, we can let $c \to 0$ in (4). As a reference, $c = 0.186$ for 802.11b. In the limit $c \to 0$, only the MIS (the right-most states in the state-transition diagram) have non-zero probabilities. Furthermore, all the MIS are equiprobable. Thus, Conditions 1, 2, and 3 lead to Propositions 1 and 2.

A justification for the approximation of ignoring the non-maximum maximal independent sets (e.g., state 0010 in the example network), even when $c$ is not very small, is that their probabilities may actually be smaller than the results of (4) after collisions have been taken into account. For the example network, link 2 has more neighbors and is more likely to experience collisions, which may result in its effective $c$, say $c_2$, being larger than those of other links after backoff is taken into account (this can be seen by making the approximation $P_{0100} = P_{0000}/c_2$).

*Mapping Normalized Throughputs to Actual Throughputs*

As a final step, we need to convert the normalized throughputs to throughputs in bps. This is the engineering part with different alternatives. This paper adopts a very simple procedure. We simply multiply the normalized throughputs by the raw throughput of a single isolated link to obtain the throughputs in bps (see section 2). We note the following: (i) this procedure may under-estimate the throughputs because in the single isolated link case, the countdown time is not shared, whereas in the multiple-link case, the countdowns of different links may occur concurrently; (ii) the BoE approximation ignores collisions and this may lead to over-estimation of throughputs. Thus, (i) and (ii) have opposing effects. Of course, more sophisticated perturbation techniques could be used to adjust for the possibilities of simultaneous countdowns and collisions. However, NS2 simulation results indicate that our simple technique is good enough for most topologies (see Fig. 2).

## 4. INSENSITIVITY TO COUNTDOWN TIME AND TRANSMISSION TIME DISTRIBUTIONS

This section removes Condition 3. The reader could also go directly to appendix B or appendix C for an alternative approach, although the treatment here is more revealing as to what is happening within the system physically. If the reader is willing to accept the insensitivity result of this section, and is more interested in the implications and applications of the results of BoE and ideal CSMA networks, he/she could proceed to section 5 directly.

In this section, we show that the formula given by (4) remains valid with either one of the following properties: (i) invariance of residual countdown-time and transmission-time distributions of a link at transition epochs of other links (in subsection 4.2); or (ii) the state probability distribution is a Markov random field (in subsection 4.3). That is, either (i) or (ii) alone will let us derive (4). These two properties are just manifestations of the fundamental behavior of the physical system rather than requirements. It is possible to prove (i), for example, by looking at a **continuous state-space** Markov process associated with the ideal CSMA model. This approach is taken in appendix B, with (i) as a corollary of the result. We have also run many Matlab simulations without explicit incorporation or artificial forcing of property (i) or (ii), and all results indicate that the formula given by (4) is valid in general.

### 4.1 Time Reversibility of Non-Markovian Processes

It is well known that time reversibility [11] is a property that can greatly simplify the analysis of stochastic processes because detailed balance [11] then applies. It is quite obvious and easy to prove that the on-off telegraph process of an isolated link is time-reversible regardless of the form of $f(t_{cd})$ and $g(t_{tr})$. From an intuitive standpoint, it is also conceivable that the interacting on-off telegraph processes of links in a network are also time-reversible. Subsection 4.4 argues that it is formally.

Most analysis of time-reversible stochastic processes in the literature focus on Markov processes. Without Condition

3, the state $S$ of our ideal CSMA network is decidedly non-Markovian. Consider the case in which all packets have the same fixed transmission time. With respect to Fig. 6, given a transition from 1000 to 1010, the next transition will be from 1010 to 0010 with probability 1, since link 1 will finish its transmission before link 3 given that it starts first. On the other hand, given a transition from 0010 to 1010, the next transition will be from 1010 to 1000. Thus, the state prior to 1010 has an impact on the future state evolution from 1010.

Nevertheless, we will show shortly that with a modified interpretation, "detailed balance" of the state probability distribution still applies to non-Markovian time-reversible systems, such as ours. For simplicity, we adopt a loose, intuitive, definition of time reversibility, as follows. The argument in subsection 4.4 that our system is time-reversible does not depend on this definition. Here, we mainly use the "converse" part of this definition for contradiction proofs: that is, if the result we want to prove is not valid, then the system is not time-reversible as defined.

**Definition of Time Reversibility:** Consider the evolution of $S(t)$ at over a long time interval $[0,T]$. Define the reverse-time trace of $S(t)$ to be $S^r(t) = S(T-t)$. If we cannot statistically distinguish between the forward-time and reverse-time traces, then the system is said to be time-reversible. Conversely, if we can tell which is which by analyzing the statistics embedded in $S(t)$ and $S^r(t)$, then the system is not time-reversible.

While the transition rates in a Markov process are well-defined, their definitions are not as clear for $S(t)$, which is non-Markov. This is so because the transition rate from one state to another in the latter is not a constant, but may depend on the prior state evolution, as shown in the above example. The following defines transition rates of non-Markov processes in terms of the "average" happenings over a long stretch of time rather than the "instantaneous" happenings.

**Definition of Transition Rate:** Suppose we conduct an experiment on the system. Consider two connected states $s$ and $s'$. Let $n_{ss'}(T)$ be the number of transitions from $s$ to $s'$ observed during a long time interval $[0,T]$. Define $n'_{ss'} = \lim_{T \to \infty} n_{ss'}(T)/T$. Let $P_s$ be the fraction of time the system is observed to be in state $s$ (i.e., $P_s = \lim_{T \to \infty} T_s / T$ where $T_s$ is the amount of time the system is in state $s$). We adopt the following definition for the transition rate: $p_{ss'} \triangleq n'_{ss'}/P_s$.

**Pre-lemma 1:** If $S(t)$ is time-reversible, $P_s p_{ss'} = P_{s'} p_{s's}$.
**Proof:** If $P_s p_{ss'} \neq P_{s'} p_{s's}$, then by definition $n'_{ss'} \neq n'_{s's}$, meaning we could statistically distinguish between $S(t)$ and $S^r(t)$ because $n'_{s's}$ is mapped to $n'_{ss'}$ in the reverse trace. □

**Comment:** The whole point of establishing this result is to show that detailed balance still applies to a time-reversible non-Markov process with the definition of $p_{ss'} \triangleq n'_{ss'}/P_s$.

**Lemma 1:** In an ideal CSMA network, $P_s p_{ss'} = P_{s'} p_{s's}$.

**Proof:** This follows from our argument in subsection 4.4 that $S(t)$ is time-reversible. □

So far $p_{ss'}$ has been defined as a quantity that can be measured from an experiment by measuring $n'_{ss'}$ and $P_s$. For all pairs of connected states $s$ and $s'$, with $s$ being the left state and $s'$ being the right state, it is not at all clear that $p_{s's}/p_{ss'} = c$. Failing that, we cannot claim that the non-Markov process has the same $P_s$ as the corresponding CTMC. Lemma 2 in subsection 4.2 shows that $p_{s's}/p_{ss'} = c$ with an additional property.

## 4.2 Invariance of Residual Countdown-Time and Transmission-Time Distributions at Transition Epochs of Other Links

Consider a link $i$ and only the sub-time intervals within $[0,T]$ during which it is actively counting down. Suppose that we randomly choose a point within these sub-time intervals for observation. The residual countdown time is the remaining countdown time before countdown reaches zero. It is well known from renewal theory that its probabability density is

$$f_{RC_i}(rc_i) = \frac{1}{E[T_{cd}]}(1 - F(rc_i))$$

where $F(rc_i) = \int_0^{rc_i} f(t_{cd})dt_{cd}$. Similarly, if we consider the times during which link $i$ is transmitting, the residual transmission time is

$$g_{RT_i}(rt_i) = \frac{1}{E[T_{tr}]}(1 - G(rt_i))$$

where $G(rt_i) = \int_0^{rt_i} g(t_{tr})dt_{tr}$. The above residual time distributions are obtained assuming all points during countdown (or transmission) are equally likely to be chosen for observation. Suppose we narrow our observation to only those points during which another link $j$ changes state from 0 to 1, or 1 to 0. We claim that that the residual time distributions remain invariant due to the randomizing effect of the system (e.g., in Fig. 6, the distribution of the residual countdown time of link 1 at transition $0000 \to 0010$, and the residual transmission time of link 1 at transition $1010 \to 1000$, are invariant). That is, the transition epochs of other links are random points within the countdown and transmission times of link $i$. After link $i$'s countdown is frozen, it may become unfrozen later when a neighbor link $j$ switches state from 1 to 0. We also claim that the distribution of the residual countdown time link $i$ is invariant at these epochs (e.g., in Fig. 6, the distribution of the residual countdown time of link 1 at transition $0100 \to 0000$). These claims have been verified by Matlab simulations under various distributions of $f(t_{cd})$ and $g(t_{tr})$, and can be proved rigorously using an approach that looks at a continuous state space associated with the system. To avoid a mouthful of words, we will refer to the above simply as the *invariant residual-time property*. Given this property, we show that $p_{s's}/p_{ss'} = c$ below.

**Pre-lemma 2:** With the invariant residual-time property, if $S(t)$ is time-reversible then $p_{s's}/p_{ss'} = c$ for all pairs of connected states $s$ and $s'$, where $s$ is the left state and $s'$ is the right state.

**Proof:** Let link $i$ be the link that is transmitting in $s'$ but not in $s$. Consider a very large time window $[0,T]$. Our proof consists of looking only at those times when the system is in either $s$ or $s'$. We make the following observations:

(i) If we conduct an experiment on the time-reversible system, $n_{ss'}(T) = n_{s's}(T) + \varepsilon(T)$ where $\varepsilon(T)$ is an error term. $\lim_{T\to\infty} \varepsilon(T)/T = 0$. This pre-lemma is about $p_{ss'}$ and $p_{s's}$, which are defined in terms of $n'_{ss'} = \lim_{T\to\infty} n_{ss'}(T)/T$ and $n'_{s's} = \lim_{T\to\infty} n_{s's}(T)/T$. Assuming the convergence of $n'_{ss'}$ and $n'_{s's}$, $\lim_{T\to\infty} \varepsilon(T)/T = 0$. Therefore, we could ignore the error term and assume that for every transition $s' \to s$, there is a reverse transition $s \to s'$ in our "thought experiment" in (ii) and (iii) below. The unmatched transitions in $\varepsilon(T)$ goes away after we divide it by $T$ and letting $T \to \infty$.

(ii) Suppose that the system makes a transition from $s'$ to $s$ at time $t$. A new random countdown time $RC_i(t^+)$ is generated for link $i$ according to $f(t_{cd})$. While in state $s$, the countdown of link $i$ proceeds, possibly together with the countdowns and transmissions of some other links. If the countdown of link $i$ finishes first, then the system moves back to state $s'$. If the countdown (or the transmission) of another link finishes first, then the system moves to another state, say state $s''$ (see Fig. 7) at time $t+\tau$. From $s''$, the system may traverse many other states before returning to either $s'$ or $s$. Say, the next time the system moves back from $s''$ to $s$ is $t+\sigma$. In general, it is possible, and quite likely, that $RC_i(t+\sigma)$ and $RC_i(t+\tau)$ are not the same, so that the countdown of link $i$ does not continue where it left off. However, with time reversibility, for every transition $s \to s''$, there is a reverse transition $s'' \to s$. With the invariant residual-time property, things are even more specific. With every transition with $RC_i = rc_i$ (more precisely, since $RC_i$ is continuous, $RC_i \in [rc_i, rc_i + drc_i)$), there must be a reverse transition from $s''$ to $s$ with $RC_i = rc_i$. This reverse transition may not correspond to the reverse transition at $t+\sigma$, but it must occur at some time earlier or later than $t+\tau$ given time reversibility and the invariant residual-time property. In particular, we could reshuffle the times within $[0,T]$ during which the system is in state $s$ so that it appears as if link $i$ is continuously counting down to zero. Each countdown-to-zero instance corresponds to a transition $s \to s'$. In particular, the system must stay in state $s$ for an average duration of $E[T_{cd}]$ for each transition $s \to s'$.

(iii) Similar argument applies for the case of a transition from $s$ to $s'$. Although afterward it is possible for the system to evolve from $s'$ to another state rather than $s$, the system must stay in state $s'$ for an average duration of $E[T_{tr}]$ for each transition from $s'$ back to $s$. Thus, $E[T_{tr}]P_s = E[T_{cd}]P_{s'} \Rightarrow p_{s's}/p_{ss'} = c$ □

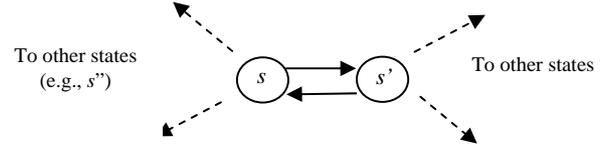

Fig. 7. Transitions between two connected states $s$ and $s'$.

**Lemma 2:** In an ideal CSMA network with the invariant residual-time property, $p_{s's}/p_{ss'} = c$ for all pairs of connected states $s$ and $s'$, where $s$ is the left state and $s'$ is the right state.

**Proof:** This follows from our argument in subsection 4.4 that an $S(t)$ ideal CSMA network is a time-reversible system. □

**Theorem 1:** The state probability distribution of an ideal CSMA network is given by (4), even if $f(t_{cd})$ and $g(t_{tr})$ are not exponentially distributed.

**Proof:** This follows from Lemmas 1 and 2 directly. □

To validate Theorem 1, we have performed Matlab simulations with Condition 1 only, without assuming exponential countdown and transmission times. Perfect fits are obtained for various distributions of countdown and transmission times, including uniform distribution for countdown time and fixed transmission time. The normalized throughputs of links and the airtime occupied by each network state are exactly as predicted by (4).

### 4.3 $P_s$ is a Markov Random Field

Markov random field is a relatively new branch of probability theory. It was originally motivated from statistical physics for the modeling of the interactions of physical entities in space. Consider a general system consisting of $L$ entities. The relationships between the entities are modeled by a graph $G$, in which vertex $i$ corresponds to entity $i$, and an edge joins two vertices if they could interact with each other. The value of the system state is a vector consisting of the values of individual states of the entities, $s = s_1 s_2 ... s_L$, with stationary probability distribution $P_s$. $P_s$ is said to be a Markov random field if $P_{s_i | s_{G-i}} = P_{s_i | s_{N_i}}$, where $s_{G-i}$ denotes the states of all other entities in the graph $G$ except $s_i$, $N_i$ denotes the neighbors of $i$, and $s_{N_i}$ denotes the states of the neighbors of $i$ only. That is, given the states of its neighbors, the state of entity $i$ is independent of the states of all other entities in the system.

It turns out that $P_s$ in our ideal CSMA network is a Markov random field given the invariant residual-time property: the "only if" part of the proof of Theorem 2 shows that the probability distribution given by (4) is a Markov random field; and Theorem 1 states that our ideal CSMA network yields (4). The "if" part is an alternative proof of Theorem 1 assuming the Markov random field property in lieu of the time-reversibility and invariant residual-time argument in the previous subsection.

**Theorem 2:** $P_s$ of an ideal CSMA network is given by (4) if and only if $P_{s_i | s_{G-i}} = P_{s_i | s_{N_i}}$ for all $i$.

**Proof:** If – Let $P_{s_i, s_{G-i}}$ denote the probability of the system state in which the state of link $i$ adopts the value of $s_i$, and the states of other links adopt the values in $s_{G-i}$. By the nature of the system, over a long stretch of time, the ratio of the airtimes used for active countdown and transmission for link $i$ must be $c$. Thus, we have $\sum_{s_{G-i}} P_{0, s_{G-i}} = c \sum_{s_{G-i}} P_{1, s_{G-i}}$, where the summation is over all $s_{G-i}$ such that the system states $0, s_{G-i}$ and $1, s_{G-i}$ are connected. Note that for each of such $s_{G-i}$, link $i$ is actively counting down in state $0, s_{G-i}$ and transmitting in state $1, s_{G-i}$. Therefore, in each of the $s_{G-i}$, $s_j = 0$ for all $j \in N_i$. Thus, we can write

$$\sum_{s_{G-i-N_i}} P_{0, \underline{0}, s_{G-i-N_i}} = c \sum_{s_{G-i-N_i}} P_{1, \underline{0}, s_{G-i-N_i}} \tag{7}$$

where the $\underline{0}$ in the indices of $P_{0, \underline{0}, s_{G-i-N_i}}$ and $P_{1, \underline{0}, s_{G-i-N_i}}$ denote the fact that $s_j = 0$ for all $j \in N_i$, and $s_{G-i-N_i}$ are the states of links who are not neighbors of $i$. If $P_{s_i | s_{G-i}} = P_{s_i | s_{N_i}}$, we have

$$\frac{P_{0, s_{G-i}}}{P_{1, s_{G-i}}} = \frac{P_{0, \underline{0}, s_{G-i-N_i}}}{P_{1, \underline{0}, s_{G-i-N_i}}} = \frac{P_{0 | \underline{0}, s_{G-i-N_i}}}{P_{1 | \underline{0}, s_{G-i-N_i}}} = \frac{P_{0 | \underline{0}}}{P_{1 | \underline{0}}} \tag{8}$$

where the $\underline{0}$ in the indices $0 | \underline{0}$ and $1 | \underline{0}$ denotes the event that $s_j = 0$ for all $j \in N_i$. Plugging (8) into (7) gives $P_{0|\underline{0}} / P_{1|\underline{0}} = c$, and re-plugging this back into (8) yields (4).

Only if – For the case where $s_j = 1$ for some $j \in N_i$, by the nature of the system, it is clear that $P_{s_i | s_{G-i}} = P_{s_i | s_{N_i}}$ because the conditional probability is zero if $s_i = 1$, and one if $s_i = 0$ regardless of $s_{G-i-N_i}$. For the case where $s_j = 0$ for all $j \in N_i$, detailed balance gives

$$P_{0, \underline{0}, s_{G-i-N_i}} = c P_{1, \underline{0}, s_{G-i-N_i}} \ \forall s_{G-i-N_i} \Rightarrow P_{0|\underline{0}, s_{G-i-N_i}} = c P_{1|\underline{0}, s_{G-i-N_i}} \ \forall s_{G-i-N_i}$$
$$\Rightarrow P_{0|\underline{0}, s_{G-i-N_i}} = c/(1+c) \ , P_{1|\underline{0}, s_{G-i-N_i}} = 1/(1+c) \ \forall s_{G-i-N_i}$$

Thus, $P_{s_i | \underline{0}, s_{G-i-N_i}}$ is independent of $s_{G-i-N_i}$, and $P_{s_i | s_{G-i}} = P_{s_i | s_{N_i}}$.
□

Although here we use the **invariant residual-time property** or the **Markov random field property** to prove that (4) is insensitive to the distributions $f(t_{cd})$ and $g(t_{tr})$, such insensitivity can be proved using an alternative approach that looks at a **continuous state space** associated with the system (see appendix B) without using either of the properties. In particular, the invariant residual-time property becomes a corollary. Then, the Markov random field property also follows as a corollary from the "only if" part of Theorem 2. In other words, both properties are intrinsic to the system and not "required assumptions".

### 4.4 Time Reversibility of $S(t)$

Physical systems are often characterized by equations. Newton's second law of motion $F = m d^2 x / dt^2$, for example, can be shown to be time-reversible by its invariance under the transformation $t \rightarrow T - t$ (i.e., by substituting $t' = T - t$ and showing $F = m d^2 x / dt'^2$ ). The reverse trace of a motion under the law is also a possible motion under the same law. Many engineering systems are not specified in terms of equations. The ideal CSMA system is defined in terms of its **operating logic**. Here, we go directly to its logic specification to demonstrate reversibility. As mentioned in subsection 3.2, the operating logic of an ideal CSMA link can be specified in terms of a simulation module, an approach taken by us here. Our argument is not to be confused with "proof by computer" in which all cases to be proved are exhaustively examined by a computer program. Rather, we perform the transformation $t \rightarrow T - t$ in the logic specification and show that the specification remains invariant. Our argument is analytical rather than simulation-based. We go through three steps in subsections 4.4.1, 4.4.2, and 4.4.3 to argue that $S(t)$ is reversible.

### 4.4.1 Forward-Time Simulation to Generate $S(t)$

With reference to Fig. 4, the following is a sketch of an *event-driven* simulation module for a link *i*. We assume there is an underlying event handler/scheduler that receives events created by links for scheduling. The current time kept by the event handler, $t_{curr}$, runs from 0 to *T*. After handling each event, the event handles the next event in its event list. In the event that link *i*'s countdown reaching zero, besides link *i*'s code segment (1) below, the event handler also freezes the countdown of link *j* if link *j* is a neighbor of link *i* and it is actively counting down (i.e., the event handler executes link *j*'s code segment (4)). In the event of link *i* beginning to count down (finishing transmission), besides link *i*'s code segment (2), the event handler also checks to see if all neighbor links *j*'s $R_j$ could be turned back to 0; and if so, executes link *j*'s code segment (3). Please refer to Fig. 4 also.

**Ideal link-model specification as a simulation module for link *i*:**
(1) Upon event "link *i* counts down to zero":
    // exits *Active Countdown State*, enters *Transmission State*
    Create event " link *i* begins to count down" to occur at time
    $$t_{next}^i = t_{curr} + t_{tr}^i \qquad (A)$$
    where $t_{tr}^i$ is a random variable generated according to $g(t_{tr}^i)$;

    // The next two lines collect the state data. $t_{prev\ trans\ ends}^i$ is the
    // epoch at which the previous transmission ends
    Set $S_i(t) = 0$ for all $t \in [t_{prev\ trans\ ends}^i, t_{curr})$ ;
    Set $S_i(t) = 1$ for all $t \in [t_{curr}, t_{next}^i)$ ;  (B)
    $t_{prev\ trans\ ends}^i = t_{next}^i$ ;

(2) Upon event "link *i* begins to count down":
    // exits *Transmission State*, enters *Active Countdown State*
    Create event "link *i* counts down to zero", **tentatively** to occur at
    $$t_{next}^i = t_{curr} + t_{cd}^i \qquad (C)$$
    where $t_{cd}^i$ is a random variable generated according to $f(t_{cd}^i)$ ;

(3) Upon event "link *i* resumes count down";
    // exits *Frozen Countdown State*, enters *Active Countdown State*
    Create event, "link *i* counts down to zero", **tentatively** to occur at
    $$t_{next}^i = t_{curr} + RC_i \qquad (D)$$
    where $RC_i$ is the remaining countdown time ;

(4) Upon event "link *i* freezes countdown":
    // enters *Frozen Countdown State*
    $RC_i = t_{next}^i - t_{curr}$ ;  (E)
    Delete the **tentative** event "link *i* counts down to zero" previously created in (2) or (3);

Note that in the above, the duration of each countdown is **tentative** because of the possibility of freezing. For each link *i*, we could either pre-generate a sequence of countdown times and transmission times to be used, or generate the times on-the-fly as we run the program: $t_{cd}^i[1], t_{tr}^i[1], t_{cd}^i[2], t_{tr}^i[2], \ldots$ .

### 4.4.2 Reverse-Time Simulation to Generate $S(t)$

Instead of the forward-time simulation, we could write an equivalent reverse-time simulation module that runs in reverse time. By reverse-time simulation, we **do not** mean generating the reverse-time $S^r(t)$ here; rather we mean using a simulation program that runs from time *T* to 0, which generates $S(t_k)$ first before $S(t_j)$ for $k > j$ for the trace $S(t)$. Provided we write the code correctly, the $S(t)$ generated as such should be statistically equivalent to the $S(t)$ generated using the forward-time simulation, because both simulations specify the same system.

The code for the reverse-time simulation is as follows. For this program, the event handler schedules events in reverse time, so that $t_{curr}$ progresses from *T* to 0; and the event list in the event handler is sorted in reverse-time order. After handling each event, the event handles the next event in its event list. In the event of link *i* beginning to count down, since we are simulating in reverse time, this event corresponds to the end of an earlier transmission. Thus, in code segment (1) below, we create an event to happen at an earlier time which corresponds to the beginning of the transmission - i.e., "link *i* counts down to zero". Besides link *i*'s code segment (1), the event handler also checks to see if the $R_j$ of each neighbor link *j* of link *i* could be turned back to 0; and if so, executes code segment (4) of link *j*. In the event of link *i*'s countdown reaching zero, besides link *i*'s code segment (2) below, the event handler also freezes the countdown of link *j* if link *j* is a neighbor of link *i* and it is actively counting down (i.e., the event handler executes link *j*'s code segment (3)) .

**Ideal link-model specification as a reverse-time simulation module:**
(1) Upon event "link *i* begins to count down":
    // exits *Transmission State*, enters *Active Countdown State*
    Create event, "link *i* counts down to zero" to occur at time
    $$t_{next}^i = t_{curr} - t_{tr}^i \qquad (A')$$
    where $t_{tr}^i$ is a random variable generated according to $g(t_{tr}^i)$ ;

    // $t_{next\ trans\ begins}^i$ is the epoch at which the next transmission begins
    Set $S_i(t) = 0$ for all $t \in [t_{curr}, t_{next\ trans\ begins}^i)$ ;
    Set $S_i(t) = 1$ for all $t \in [t_{next}^i, t_{curr})$ ;  (B')
    $t_{next\ trans\ beginss}^i = t_{next}^i$ ;

(2) Upon event "link *i* counts down to zero":
    // exits *Active Countdown State*, enters *Transmission State*
    Create event "link *i* begins to count down", **tentatively** to occur at
    $$t_{next}^i = t_{curr} - t_{cd}^i \qquad (C')$$
    where $t_{cd}^i$ is a random variable generated according to $f(t_{cd}^i)$ ;

(3) Upon event "link *i* freezes countdown":
    // exits *Active Countdown State*, enters *Frozen Countdown State*
    Create event "link *i* begins to countdown", **tentatively** to occur at

$$t_{next}^i = t_{curr} - RC_i \qquad (D')$$

where $RC_i$ is the remaining countdown time ;

(4) Upon event "link $i$ resumes countdown"
//exits *Frozen Countdown State* :

$$RC_i = t_{curr} - t_{next}^i ; \qquad (E')$$

Delete the **tentative** event "link $i$ begins to count down" previously created in (2) or (3);

If we use the end state at $T$ found in a forward-time simulation as the initial state here (including the remaining count-down times and the remaining transmission times of all links), and then use the same sequence of countdown and transmissions times $t_{cd}^i[1], t_{tr}^i[1], t_{cd}^i[2], t_{tr}^i[2], ... \forall i$ (but in reverse), the $S(t)$ produced should be identical (not just statistically equivalent) to that of the forward-time simulation.

If the system is stationary in that the limiting probability $\lim_{T \to \infty} P_{S(T)}(s)$ exists, then the initial states for forward-time and reverse-time simulations do not matter, and that the $S(t)$ obtained from forward-time and reverse-time simulations will be statistically equivalent.

### 4.4.3 Reversing $S(t)$ to Obtain $S^r(t)$

Suppose that after generating $S(t)$ using the reverse-time simulation, we reverse the trace. Specifically, we look at $S^r(t) \triangleq S(T-t)$. We claim that $S^r(t)$ and $S(t)$ are statistically equivalent, and therefore the system is time-reversible. To see this, suppose that rather than generating $S(t)$ and then do the transformation $t \to T-t$ to get $S^r(t)$, we incorporate the transformation directly into (A') to (E') to obtain $S^r(t)$ in the reverse-time simulation. To do so, we do the following in modifications in the code: $t \to T-t$ (in the event handler), $t_{next}^i \to T - t_{next}^i$, $t_{curr} \to T - t_{curr}$, $t_{next}^i \to T - t_{next}^i$, $t_{next\ trans\ begins}^i \to T - t_{next\ trans\ begins}^i$. Then, we run the simulation from $t=0$ to $t=T$. Note that $t_{tr}^i$, $t_{cd}^i$ and $RC_i$ are not time epochs, but positive time intervals that should not undergo any transformation. In our code, we then have

$$t_{next}^i = t_{curr} + t_{tr}^i \qquad (A'')$$

$$t_{next}^i = t_{curr} + t_{cd}^i \qquad (C'')$$

$$t_{next}^i = t_{curr} + RC_i \qquad (D'')$$

$$RC_i = t_{next}^i - t_{curr} \qquad (E'')$$

After the above transformation, when we gather data, time has already been reversed. The correct modification of $S(t)$ for (B') is

set $S_i^r(t) = 0$ for all $t \in [t_{next\ trans\ begins}^i, t_{curr})$

set $S_i^r(t) = 1$ for all $t \in [t_{curr}, t_{next}^i) \qquad (B'')$

$t_{next\ trans\ begins}^i = t_{next}^i$ ;

We note that the new code is identical to the forward-time simulation code in subsection 4.2.1 except that variable $t_{next\ trans\ begins}^i$ here is $t_{prev\ trans\ ends}^i$ there; event "link $i$ begins to count down" here is event "link $i$ counts down to zero" there; etc. But these are just variable names and labels that would not affect the $S^r(t)$ produced. If we want, for clarity we could relabel $t_{prev\ trans\ ends}^i$ here as $t_{next\ trans\ begins}^i$; event "link $i$ begins to count down" here as event "link $i$ counts down to zero" ; etc. (note that looking at the trace in reverse means "counts down to zero" is "begins count down", etc ). Given the code for generating $S^r(t)$ is identical to the code for generating $S(t)$, $S^r(t)$ and $S(t)$ should therefore be statistically identical.

## 5. IMPLICATIONS AND APPLICATIONS OF BoE

So far we have focused on explaining BoE. In this section, we briefly discuss the implications and applications of BoE results.

*Global optimality and local starvation/unfairness* – The value of $c$ does not have to be extremely small for BoE approximation to be good. For example, in 802.11b networks, where $c = 0.186$, our simulation results match very well with the BoE computed results. When that happens, the system spends almost all its time among the MIS. This implies that the "greedy" CSMA distributed protocol achieves the **highest global throughpu**t, since the number of simultaneously transmitting links are maximum in MIS. On the other hand, starvation of specific links is a common phenomenon (e.g., see results in Fig. 2). Indeed an application of BoE is to quickly identify starved links so that remedial actions could be taken. The remedy could be to assign the starved links to other frequency channels, or to make $c$ non-uniform among links. For the latter, consider the network of Fig. 1 in which link 2 is starved. To unstarved it, one could change its $c$, either by reducing the average countdown time, or the average transmission duration (i.e., CW and TXOP, respectively, in the parlance of 802.11 [10]). Reducing $c_2$ causes the non-MIS state 0100 to have non-negligible probability, and detailed-balance analysis similar to that of (4) can be used to set $c_2$. The reader could verify that with $c_1 = 1, c_2 = 0.012, c_3 = c_4 = 0.024$, the links would have roughly equal normalized throughputs of 0.33.

*Resource allocation* – In a general wireless network, each independent set represents the links that can transmit simultaneously without detrimental effects. When considering resource allocation in a **centrally controlled** wireless network, we could assign a weight to each independent set to correspond to the amount of airtime allocated to that independent set; the links in the independent set then transmit together during the allocated airtime. In assigning the weights, the utility to be optimized could be the total system throughput, proportional fairness utility, max min throughput, etc. [13], etc. The optimal solution is often computed in a **centralized** manner in which global knowledge of the contention graph is assumed. The CSMA protocol considered by this paper, on the other hand, is a **distributed resource allocation** protocol. To the extent that $c$ is small, BoE of this paper states that only the MIS among all the independent sets are allocated appreciable airtime, and that all the MIS are allocated equal airtime. More generally, the utility to be achieved under CSMA can be modified by adjusting $c_i$, $i=1,...,L$. For example, our goal could be to maximize the total system throughput, $\sum_i Th_i$, subject to certain ratios of throughput requirements, $Th_i / Th_1 = r_i \ \forall \ i = 2,...,L$. An insight of this paper is that under the ideal CSMA system the form of the distributions of countdown and transmission times are immaterial and cannot be used for resource allocation purposes once the ratio of their means is given. We only have $L$ degrees of freedom for our resource allocation problem, because the variables of optimization are $c_i$, $i=1,...,L$. This is in contrast to the general weight-assignment-to-independent-set problem mentioned above in which there are usually more than $L$ weights to be assigned. Indeed, if we consider the fact that we may require $c_i \geq c_0$ for some $c_0$ due to implementation limit on how small the countdown timeslot can be, we only have $L-1$ degrees of freedom left in this problem, since $\min_i c_i = c_0$ if our goal is to maximize $\sum_i Th_i$.

*Small change in topology leads to large change in throughput distribution* – It is difficult to estimate the throughput of a link based the "local" topology around it. Another link faraway can affect it significantly. Consider Fig. 2. Link 2 is starved in topology 3. Adding a link 5 in topology 3 gives us topology 4, and link 2 becomes "unstarved". In topology 5, all five links in the ring have the same throughput. The addition of link 6 in topology 6 causes three links to be starved and three links to grab the maximum normalized throughput. It has not escaped our attention that identifying the MIS in a graph is an NP-complete problem, and as such BoE can be complex when the graph is large. However, in view of the fact that a small change in a graph can trigger large changes in the set of MIS (hence the throughput distributions), it is doubtful that this complexity can be avoided: heuristics that make incremental adjustments to computed throughputs based on incremental topological perturbations are not viable.

Fortunately, for modest-size networks, BoE computation is indeed very quick.

*Island States* – The throughput distributions computed by BoE are the long-term averages. Temporal starvation may still occur even though the long-term average throughput is acceptable. To see this, consider topology 7 in Fig.2. The MIS are 101010 and 010101. According to BoE, the long-term normalized throughput of each link is 0.5. However, the Hamming distance between these two MIS is 6, meaning the states of six links have to change in order to move from one MIS to the other MIS. When $c$ is small and the Hamming distance between two MIS is large, such a move occurs only rarely, as if the two MIS are islands separated by oceans. In the network, three of the links can be starved for a long duration of time. Temporal starvation, and the existence of island states, can be identified from the state-transition diagram and Hamming distance analysis. Such "long-term oscillatory behavior" has also been observed in [14] in its TCP over 802.11 wireless network simulation results. Our work suggests that even without TCP, CSMA networks still exhibit such behavior.

*Networks with hidden nodes* – This paper has focused on networks without hidden nodes. A network without hidden nodes can be designed based on the principle of hidden-node free design in [15]. Whether the analytical approach in this paper can be extended for networks with hidden nodes is an interesting subject for further studies.

## 6. CONCLUSIONS

We have presented a simple BoE method for computing throughput distributions among links in a CSMA network. For 802.11 networks, this fast computation method has been verified to be very accurate.

We have also developed a theory to explain BoE. Besides explaining BoE, the theory also reveals some rather interesting mathematical properties of processes that are **non-Markovian in time** associated with physical systems consisting of entities interacting in **time and space**.

This paper models an **ideal CSMA network** as a set of **interacting on-off telegraph processes**. In this model, a link is in the "on" state if it is transmitting; in the "off" state otherwise; and the duration of the "off" state is affected by the neighbor links by means of the freezing of its residual time when a neighbor on-off process is in the "on" state. The **system state**, defined by the set of "on" processes, is in general non-Markov under general distributions of the "on" and "off" durations. This is possibly the first paper to prove the mathematical result that the state probability distribution $P_s$ is insensitive to the distributions of the "on" and "off" durations given the ratio of their means. In addition, we have shown that the system state is **time-** **reversible**. Furthermore, $P_s$ is a **Markov Random Field** (MRF) [8, 11].

MRF was originally motivated from statistical physics. In engineering, it has found applications in image analysis and computer vision. In addition, it has been investigated for sensor-data analysis and *ad hoc* network routing [16]. However, most of the engineering applications are in the context of signal processing and decision making. As far as we know, this paper establishes for the first time that the interference and carrier-sensing interactions in a wireless network result in a random process that is an MRF. The fact that this has eluded researchers in the field is rather surprising, considering that contention-graph modeling, a popular technique in wireless research, finds a perfect fit in the definition of MRF. The techniques and results of MRF may have an important role to play for work beyond this paper.

## APPENDIX A: PRIOR METHODS IN COMPARISON WITH BOE

In this appendix, we discuss the difficulty of using previous methods to compute throughput distributions in wireless networks in which the carrier-sensing relationships among the links are not all inclusive. We first consider the approach in [1].

Ref. [1] considers *homogeneous* networks in which all links experience the same situation. This allows one to analyze the state transition of one link to deduce the link throughput, from which the overall network throughput is then obtained by multiplying the link throughput by the number of links. The state transition of a link $i$ can be represented by a Markov chain in which the state is $(Co_i, RC_i)$, where $Co_i$ is the number of collisions previously experienced by the packet of the link, and $RC_i$ is the remaining number of countdown timeslots before the next transmission attempt. Both $Co_i$ and $RC_i$ are discrete variables.

In the homogeneous case, all links count down at the same time. When one link transmits, countdowns of all links also freeze at the same time due to all-inclusive carrier sensing. This is not the case in a non-all-inclusive carrier-sense network, such as that in Fig. 1, in which link 2 will freeze when link 1 transmits, but links 3 and 4 will continue to count down. The state transition and evolution of the overall network is highly dependent on the associated contention graph and can be quite complicated. In particular, due to the inhomogeneity, it is not enough just to represent the state of a link with $(Co_i, RC_i)$. The transmission state of the link will also need to be captured, because it affects different links differently. In general, we need to specify $(Co_i, S_i, RC_i, RT_i)$, where $S_i = 1$ if link $i$ is transmitting, and $S_i = 0$ otherwise; and $RT_i$ is the remaining transmission time if $S_i = 1$. The total number of states for the overall network can be quite prohibitive. The transitions among the states are highly dependent on the associated contention graph. Trying to derive the throughputs of links in a general way is rather formidable using this approach.

In justifying BoE, our paper here takes a simplifying approach that treat $RC_i$ and $RT_i$ as continuous variables rather than discrete variables, and in doing so, not we only reduce the state space to be considered, but also allow collisions to be ignored in the first-cut analysis. The assumption is that the effects of collisions can be taken into account later by a perturbation adjustment, **if necessary**. It turns out that no sophisticated perturbation technique is needed for 802.11 networks, as can be seen in the procedure of BoE specified in section 2. In particular, BoE reduces the possible states of each link to only 2 (although our analytical justification for this simplicity presented in sections 3 and 4 cannot pre-assume this a priori).

Next, we consider the approximate method in [3]. The method in [4] is similar except that packet collisions are integrated into the analysis. To be fair, [2-4] considered the multi-hop scenario rather than the single-hop scenario (focus of this paper); here we just apply the method to the special case of the single-hop scenario. The crux of the method is to examine the local observations as experienced by individual nodes. Consider a long stretch of time interval in $[0,T]$. A node can only observe the airtimes used by the nodes within its carrier-sensing range. Let $S_i$ and $C_i$ be the airtimes within $[0,T]$ that are used by a "steady-state" node $i$ to transmit and count down, respectively (note: the notation $S_i$ is used differently here then other parts of this paper: we adopt this notation from [2, 3] for the rest of this appendix). In Fig. 1, as far as link 2 is concerned, it will observe $C_2$, $S_1$, $S_3$ and $S_4$. Note that $S_1$ may overlap with $S_3$ and $S_4$. From link 2's point of view, the total airtimes used up these nodes can not exceed $T$. With the assumption that the network is saturated, we have $|C_2 \cup S_2 \cup S_1 \cup S_3 \cup S_4| = T$, which can be further decomposed using the inclusion-exclusion principle:

$$|C_2 \cup S_2 \cup S_1 \cup S_3 \cup S_4| = |C_2| + |S_2| + \cdots + |S_4| - |C_2 \cap S_2| - |S_2 \cap S_1| - \cdots - |S_3 \cap S_4| + |C_2 \cap S_2 \cap S_1| + |S_2 \cap S_1 \cap S_3| + \cdots$$

However, it is easy to see that for the particular network, the intersections of the airtimes used by two or more nodes are null by virtue of carrier sensing (if collisions are ignored), except $|S_1 \cap S_3|$ and $|S_1 \cap S_4|$. Consider the overlapped airtimes of link 1 and link 3. When link 2 is transmitting, links 1 and 3 do not transmit due to carrier sensing. Let $x_i = |S_i|/T$ be the fraction of airtime within [0, $T$] that is used by link $i$. The remaining fraction of airtime where $S_1$ and $S_3$ may overlap is $1-x_2$. Ref. [3] makes the following approximation:

$$|S_1 \cap S_3| = \frac{x_1 x_3}{1-x_2}, \quad |S_1 \cap S_4| = \frac{x_1 x_4}{1-x_2}$$

Hence, we have

$$cx_2 + x_2 + x_1 + x_3 + x_4 - \frac{x_1 x_3}{1-x_2} - \frac{x_1 x_4}{1-x_2} = 1 \quad (A1)$$

Applying similar method on links 1, 3, and 4 gives

$$cx_1 + x_1 + x_2 = 1 \quad (A2)$$
$$cx_3 + x_3 + x_2 + x_4 = 1 \quad (A3)$$
$$cx_4 + x_4 + x_2 + x_3 = 1 \quad (A4)$$

Combing (A1) – (A4), the normalized throughput of each link is found to be (0.93, 0.08, 0.51, 0.51). Benchmarked against NS2's results, BoE is more accurate. Fig. A1 gives two more examples to show the better accuracy of BOE. In particular, the method in [3] fails to give a correct solution for the second topology.

| Contention graph | BoE | Ng and Liew [3] | NS2 |
|---|---|---|---|
| 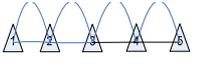 | (1, 0, 1, 0,1) | (0.82, 0.21, 0.73, 0.21, 0.82) | (0.96, 0.05, 0.93, 0.05, 0.96) |
| 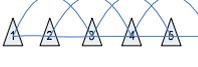 | (0.5, 0.33, 0.17, 0.17, 0.33, 0.5) | (0.52, 0.25, 0.33, 0.33, 0.25, 0.52) | (0.53, 0.37, 0.17, 0.17, 0.37, 0.53) |

Fig. A1. Comparison of BoE and Ng&Liew [3].

## APPENDIX B: INSENSITIVITY PROOF USING CONTINUOUS STATE-STATE APPROACH

This appendix provides an alternative proof that $P_s$ as expressed in (4) is insensitive to the form of $f(t_{cd})$ and $g(t_{tr})$ given their means. Recall that $S = S_1 S_2 ... S_L$ is not a Markov process in time. However, non-Markov processes can often be turned into Markov processes by changing the state definition. In fact, if we include all the dependencies of the past into our state definition, the non-Markov process becomes a Markov process. For the ideal CSMA system, let us define the state of a link $i$ as $X_i = (S_i, RC_i, RT_i)$. The **state** of the overall system is then $X = X_1 X_2 ... X_L$. We will refer to the state used in the main body of this paper $S$ as the **transmission state** here. It is interesting to note that $S(t)$ is non-Markovian but reversible in time; while $X(t)$ is Markovian but not reversible in time (because $RC_i(t)$ and $RT(t)$ are not reversible).

For a particular realization of $X$, $x$, let $s$ be the corresponding realization of $S$. Similarly, let $rc_i$ and $rt_i$ be the values of $RC_i$ and $RT_i$, respectively, under $x$. Let $\chi$ be the set of all feasible $x$.

**Theorem B1:** The stationary probability density of $X$ is

$$p_X(x) = P_s \prod_{i \in T(s)} g_{RT_i}(rt_i) \prod_{i \in C(s)} f_{RC_i}(rc_i) \prod_{i \in G-T(s)-C(s)} f_{RC_i}(rc_i)$$

$$\forall\, x \in \chi \quad (B1)$$

where $P_s$ is given by (4), $f_{RC_i}(rc_i) = \frac{1}{E[T_{cd}]}(1-F(rc_i))$, $g_{RT_i}(rt_i) = \frac{1}{E[T_{tr}]}(1-G(rt_i))$, $T(s)$ is the set of transmitting links under $s$, $C(s)$ is the set of active countdown links under $s$, and $G$ is the set of all links in the system.

**Comment**: By integrating $p_X(x)$ in (B1) over all possible values of $rt_i$ and $rc_i$ for all $i$, we get $P_s$ in (4). In other words, if (B1) is valid, the transmission-state distribution $P_s$ is insensitive to the forms of $f(t_{cd})$ and $g(t_{cd})$. Also, a **corollary** of (B1) is the **invariant residual-time distributions** mentioned in the main body of this paper: according to (B1), the remaining countdown and transmission times of different links are independent, and therefore the fact that a link is counting down to zero or completing its transmission, thus experiencing a transition, has no bearing on the residual countdown and transmission times of other links.

**Proof:** The balance equation of our system is given by (B2), and we show that (B1) satisfies the balance equation. The derivation of the balance equation is presented in the section immediately after this proof.

$$-\sum_{i \in C(s)} \frac{\partial p_X(x)}{\partial rc_i} - \sum_{i \in T(s)} \frac{\partial p_X(x)}{\partial rt_i}$$
$$= \sum_{i \in C(s)} p_X(S_{i1} RT_{i0^+} x) f(rc_i) + \sum_{i \in T(s)} p_X(S_{i0} RC_{i0^+} x) g(rt_i) \quad (B2)$$

where $RT_{i0^+}$ is the operator that sets $rt_i$ in $x$ to $0^+$ (i.e., just before transmission of link $i$ completes); $RC_{i0^+}$ is the operator that sets $rc_i$ in $x$ to $0^+$ (i.e., just before countdown of link $i$ completes); $S_{i1}$ and $S_{i0}$ are the operators that set $s_i$ in $x$ to 1 and 0, respectively.

We now show that (B1) satisfies (B2). In fact, the terms in the LHS and RHS of (B2) are matched on a one by one basis under (B1). We show that

$$-\frac{\partial p_X(x)}{\partial rc_j} = p_X(S_{j1}RT_{j0^+}x)f(rc_j) \text{ for each } j \in C(s) \text{ under}$$

(B1). The argument for

$$-\frac{\partial p_X(x)}{\partial rt_j} = p_X(S_{j0}RC_{j0^+}x)g(rt_j) \text{ for each } j \in T(s) \text{ is}$$

similar. Under (B1),

$$-\frac{\partial p_X(x)}{\partial rc_j}$$

$$= -\frac{df_{RC_j}(rc_j)}{drc_j}P_s \prod_{i \in T(s)} g_{RT_i}(rt_i) \prod_{i \in C(s)-j} f_{RC_i}(rc_i) \prod_{i \in G-T(s)-C(s)} f_{RC_i}(rc_i)$$

$$= \frac{f(rc_j)}{E[T_{cd}]} \frac{B}{c^{|T(s)|}} \prod_{i \in T(s)} g_{RT_i}(rt_i) \prod_{i \in C(s)-j} f_{RC_i}(rc_i) \prod_{i \in G-T(s)-C(s)} f_{RC_i}(rc_i)$$

(B3)

Under (B1),

$$p_X(S_{j1}RT_{j0^+}x)f(rc_j)$$

$$= f(rc_j)\frac{B \cdot g_{RT_j}(0^+)}{c^{|T(s)|+1}} \prod_{i \in T(s)} g_{RT_i}(rt_i) \prod_{i \in C(s)-j} f_{RC_i}(rc_i) \prod_{i \in G-T(s)-C(s)} f_{RC_i}(rc_i)$$

$$= \frac{f(rc_j)}{E[T_{tr}]} \frac{B}{c^{|T(s)|+1}} \prod_{i \in T(s)} g_{RT_i}(rt_i) \prod_{i \in C(s)-j} f_{RC_i}(rc_i) \prod_{i \in G-T(s)-C(s)} f_{RC_i}(rc_i)$$

(B4)

Thus, (B3) and (B4) are equal. □

**Derivation of Balance Equation (B2)**

The method we use is similar to that found in [17] for queuing network analysis. More generally, the stochastic process corresponding to our ideal CSMA system belongs to the class of piecewise-deterministic Markov processes [18].

Let $p_X(t,x)$ be the state probability density at time $t$. At equilibrium,

$$\frac{dp_X(t,x)}{dt} = \lim_{\Delta t \to \infty} \frac{p_X(t+\Delta t,x) - p_X(t,x)}{\Delta t} = 0 \quad (B5)$$

In the time interval from $t$ to $t+\Delta t$, the state changes as a result of links counting down and transmitting. It is possible that in another state $x'$, a link counts down to zero within the time interval, and as a result the state jumps to $x$. It is also possible that a link finishes transmission in another state within the time interval, and as a result the state jumps to $x$. For small $\Delta t$, the probability of having more than one such "jump event" is in order $o(\Delta t)$. There is also the possibility that links are counting down and transmitting within the time interval without any jump events being incurred, in which case $x$ changes without $s$ being changed. For a particular state realization $x$, we can write

$$p_X(t+\Delta t, x) = RT0 + RC0 + CDTR + o(\Delta t) \quad (B6)$$

where RT0 is the contribution due to end-of-transmission jump events, RC0 is the contribution due to countdown-to-zero jump events, CDTR is the contribution due to ordinary counting down and transmission without any jump events, and $\lim_{\Delta t \to \infty} o(\Delta t)/\Delta t = 0$. With the notation in the proof of Theorem B1, we can write

$$\begin{aligned} RT0 &= \sum_{i \in C(s)} p_X(t, S_{i1}RT_{i0^+}x) \cdot \Delta rt_i \cdot f(rc_i) \\ &= \sum_{i \in C(s)} p_X(t, S_{i1}RT_{i0^+}x) f(rc_i) \Delta t \end{aligned} \quad (B7)$$

Note that in the above, for link $i$ to just finish transmission within $\Delta t$, $rt_i$ must fall within the interval $(0, \Delta rt_i)$ at time $t$, where $\Delta rt_i = |drt_i/dt|\Delta t = \Delta t$. Similarly, we have

$$\begin{aligned} RC0 &= \sum_{i \in T(s)} p_X(S_{i0}RC_{i0^+}x) \cdot \Delta rc_i \cdot g(rt_i) \\ &= \sum_{i \in T(s)} p_X(S_{i0}RC_{i0^+}x) g(rt_i) \Delta t \end{aligned} \quad (B8)$$

For CDTR, in order to evolve to state $x$ at time $t+\Delta t$, each link $i$ that is actively counting down must have $RC_i = rc_i + \Delta t$, and each link that is transmitting must have $RT_i = rt_i + \Delta t$, at time $t$. That is, $\Delta rc_i = \Delta t$ and $\Delta rt_i = \Delta t$, respectively. At the risk of notational abuse, let us denote the state at time $t$ by $x + \Delta x$. By Taylor expansion, we have

$$\begin{aligned} CDTR &= p_X(t, x+\Delta x) \\ &= p_X(t,x) + \sum_{i \in C(s)} \frac{\partial p_X(x)}{\partial rc_i}\Delta t + \sum_{i \in T(s)} \frac{\partial p_X(x)}{\partial rt_i}\Delta t + o(\Delta t) \end{aligned} \quad (B9)$$

Putting (B7), (B8), and (B9) into (B6), and then taking the derivative limit in (B5), we get the balance equation (B2).

## APPENDIX C: INSENSITIVITY PROOF USING "MIXTURE OF GAMMA DISTRIBTUTIONS" APPROACH

The appendix presents the "mixture-of-gamma-distributions" limiting approach [19] to prove the insensitivity of $P_s$ given by (4) to the distributions of $f(t_{cd})$ and $g(t_{cd})$.

**Gamma Distributed** $f(t_{cd})$ **and** $g(t_{cd})$

We first consider the case where the countdown time and transmission time are gamma-distributed. Each countdown time consists of $y$ stages of exponentially-distributed constituent countdown times of mean $d \triangleq E[T_{cd}]/y$; and each transmission time consists of $z$ stages of exponentially-distributed constituent transmission times of mean $e \triangleq E[T_{tr}]/z$. Redefine $rc_i \in \{1,2,...,y\}$ as the remaining "countdown stage" of link $i$ while it is counting down, and $rt_i \in \{1,2,...,z\}$ as the remaining "transmission stage" of link $i$ while it is transmitting. Define the state of link $i$ as $X_i = (S_i, RC_i, RT_i)$, and the state of the overall system as $X = X_1 X_2 ... X_L$. We can see that $X$ is a continuous-time Markov chain (CTMC). Unless otherwise specified, other notations are similar to that in appendix B.

**Theorem C1:** Under gamma-distributed $f(t_{cd})$ and $g(t_{tr})$, the stationary probability distribution of $X$ is

$$p_X(x) = P_s \cdot \frac{1}{z^{|T(s)|}} \cdot \frac{1}{y^{L-|T(s)|}} = \frac{B}{y^L}\left(\frac{e}{d}\right)^{|T(s)|} \quad \forall \, x \in \chi \quad (C1)$$

where $P_s$ and $B$ are given by (4).

**Comment**: By integrating $p_X(x)$ in (C1) over all possible values of $rt_i$ and $rc_i$ for all $i$, we get $P_s$ in (4). In other words, if (C1) is valid, the transmission-state distribution $P_s$ remains the same as that of exponential distributions under gamma distributions for countdown and transmission times.

**Proof:** The proof makes use of Lemma C1 below. We first derive $q(x, x')$. Then we show that $P_X(x)$ in (C1) together with a "guess" for $q^r(x, x')$ satisfy (C2) and (C3) in Lemma C1. For a state $x$ and another state $x'$ "connected" to it, one of the following three cases must apply:

(i) $x' = S_{i0} RT_{i0} RC_{iy} x$ for some $i \in T(s)$ where $RC_{iy}$ is the operator that sets $rc_i$ in $x$ to $y$, $RT_{i0}$ is the operator that sets $rt_i$ in $x$ to 0, and $S_{i0}$ is the operator that sets $s$ in $x$ to 0. This is the case in which transmission of link $i$ just finishes in state $x$ and the state jumps to $x'$. Note that $rt_i = 1$ in $x$ in this case. By virtue of the exponential distribution of the last stage of the transmission, $q(x, x') = 1/e$.

(ii) $x' = S_{i1} RC_{i0} RT_{iz} x$ for some $i \in C(s)$ where $RT_{iz}$ is the operator that sets $rt_i$ in $x$ to $z$, $RC_{i0}$ is the operator that sets $rc_i$ in $x$ to 0, and $S_{i1}$ is the operator that sets $s$ in $x$ to 1. This is the case in which the countdown of link $i$ just finishes in state $x$ and the state jumps to $x'$. Note that $rc_i = 1$ in $x$ in this case. By virtue of the exponential distribution of the last stage of the countdown, $q(x, x') = 1/d$.

(iii) $x' = RC_{i, rc_i - 1} x$ for some $i \in C(s)$, or $x' = RT_{i, rt_i - 1} x$ for some $i \in T(s)$, where $RC_{i, rc_i - 1}$ is the operator that decrements $rc_i$ by one, and $RT_{i, rt_i - 1}$ is the operator that decrements $rt_i$ by one. This is the case in which $s' = s$, and either the countdown stage of some link $i$ advances by one, or the transmission stage of some link $i$ advances by one. In the former, $rc_i \geq 2$, and $q(x, x') = 1/d$. In the latter, $rt_i \geq 2$, and $q(x, x') = 1/e$.

As an illustration, consider the network shown in Fig. 1, and assume $y = z = 2$. Fig. C1(a) below shows the transitions associated with the state $x = (1,0,1)(0,0,1)(0,0,2)(0,0,1)$. That is, link 1 is in its last stage of transmission, links 2 and 4 are in their last stage of countdown (but the countdown of link 2 is being frozen), and link 3 is in its first stage of countdown. In the figure, for economy of presentation, the state is represented by two rows of numbers. The upper row is simple $s = s_1...s_L$; the element $i$ of second row is $rc_i$ if $s_i = 0$, and is $rt_i$ if $s_i = 1$.

Our "guess" for $q^r(x', x)$ for cases (i) to (iii) above is as follows:

(i) $q^r(x', x) = 1/d$

(ii) $q^r(x', x) = 1/e$

(iii) $q^r(x', x) = 1/d$ if $x' = RC_{i, rc_i - 1} x$ for some $i \in C(s)$;
$q^r(x', x) = 1/e$ if $x' = RT_{i, rt_i - 1} x$ for some $i \in T(s)$

For illustration, Fig. C1(b) shows the transitions of state $x = (1,0,1)(0,0,1)(0,0,2)(0,0,1)$ of the reverse process. Note that the process $X(t)$ is not time-reversible: the Markov chain of the reverse process in Fig. C1(b) is clearly different from that of $X(t)$ in Fig. C1(a).

It is straightforward to verify that for cases (i) to (iii), $p_X(x)$ in (C1) and $q^r(x', x)$ proposed above satisfy (C3) in Lemma C1, as follows:

(i) $p_X(x) q(x, x') = \frac{B}{y^L}\left(\frac{e}{d}\right)^{|T(s)|} \cdot \frac{1}{e}$, while

$p_X(x') q^r(x', x) = \frac{B}{y^L}\left(\frac{e}{d}\right)^{|T(s)|-1} \cdot \frac{1}{d}$. Thus, they are equal.

(ii) $p_X(x) q(x, x') = \frac{B}{y^L}\left(\frac{e}{d}\right)^{|T(s)|} \cdot \frac{1}{d}$, while

$p_X(x') q^r(x', x) = \frac{B}{y^L}\left(\frac{e}{d}\right)^{|T(s)|+1} \cdot \frac{1}{e}$. Thus, they are equal.

(iii) If $x' = RC_{i, rc_i - 1} x$ for some $i \in C(s)$, then

$p_X(x) q(x, x') = p_X(x') q^r(x', x) = \frac{B}{y^L}\left(\frac{e}{d}\right)^{|T(s)|} \cdot \frac{1}{d}$.

If $x' = RT_{i,rt_i-1}x$ for some $i \in T(s)$, then

$$p_X(x)q(x,x') = p_X(x')q^r(x',x) = \frac{B}{y^L}\left(\frac{e}{d}\right)^{|T(s)|} \cdot \frac{1}{e}.$$

It remains to show that (C2) in Lemma C1 is also satisfied. This is rather obvious. Specifically, $q(x) = |C(s)|/d + |T(s)|/e = q^r(x)$. Note that for the forward process, while in state $x$, $|C(s)|$ is the number of links actively counting down and $|T(s)|$ is the number of links transmitting. For the reverse process, we reverse count down (i.e., count up) and reverse transmit. When the count-up of link $i$ completes $y$ stages, link $i$ goes into reverse transmission. In any case, $|C(s)|$ is the number of links actively counting up and $|T(s)|$ is the number of links reverse-transmitting. □

**Lemma C1:** Denote the transition rates of the Markov process $X(t)$ by $q(x,x')$, $x,x' \in \chi$. If we can find a collection of numbers $q^r(x,x')$, $x,x' \in \chi$, such that

$$q(x) \triangleq \sum_{x' \in \chi} q(x,x') = \sum_{x' \in \chi} q^r(x,x') \triangleq q^r(x) \quad \forall\, x \in \chi \quad (C2)$$

and a collection of positive numbers $p_X(x), x \in X$ summing to unity, such that

$$p_X(x)q(x,x') = p_X(x')q^r(x'x), \quad \forall\, x,x' \in \chi \quad (C3)$$

then $q^r(x,x')$, $x,x' \in \chi$, are the transition rates of the reversed process $X(T-t)$ and $p_X(x), x \in \chi$, is the equilibrium distribution of both processes.

**Proof:** This is a general property of CTMC. Lemma C1 is just a paraphrase of Theorem 1.13 in [11]. The reader is referred to [11] for the proof. □

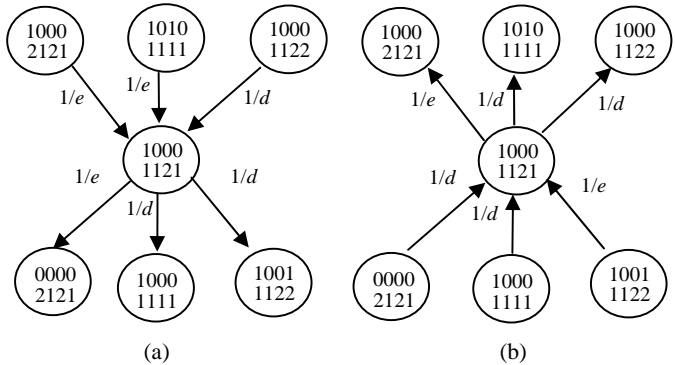

Fig. C1. Transitions of state $\frac{1000}{1121}$ for network of Fig. 1 of our paper under (a) forward process $X(t)$; (b) reverse process $X(T-t)$.

**Mixtures of Gamma Distributions**

Suppose now that $f(t_{cd})$ and $g(t_{tr})$ are mixtures of gamma distributions. Specifically, suppose that $y$ and $z$ are random, and $Y$ and $Z$ are the corresponding random variables. A new countdown time $y$ is generated with probability $P_Y(y)$; a new transmission time is generated with probability $P_Z(z)$. We require that $\sum_y d \cdot y P_Y(y) = E[T_{cd}]$ and $\sum_z e \cdot z P_Z(z) = E[T_{tr}]$. This means $c = \frac{E[T_{cd}]}{E[T_{tr}]} = \frac{d \cdot E[Y]}{e \cdot E[Z]}$. We define the state of link $i$ as $X_i = (S_i,(Y_i,RC_i),(Z_i,RT_i))$ where $Y_i$ contains the value of $y_i$ drawn when countdown started (if the link is counting down), and $Z_i$ contains the value of $z_i$ drawn when the transmission started (if the link is transmitting).

**Theorem C2:** When $f(t_{cd})$ and $g(t_{tr})$ are mixtures of gamma distribution as described above, the stationary probability distribution of $X$ is

$$p_X(x)$$
$$= P_s \frac{1}{E[Z]^{|T(s)|}} \frac{1}{E[Y]^{L-|T(s)|}} \prod_{i \in T(s)} P_Z(z_i) \prod_{i \in G-T(s)} P_Y(y_i) \quad (C4)$$
$$= \frac{B}{E[Y]^L}\left(\frac{e}{d}\right)^{|T(s)|} \prod_{i \in T(s)} P_Z(z_i) \prod_{i \in G-T(s)} P_Y(y_i) \quad \forall\, x \in \chi$$

where $P_s$ and $B$ are given by (4).

**Comment**: By summing $p_X(x)$ in (C4) over all possible values of $y_i$ and $z_i$ and over all possible values of $rc_i$ and $rt_i$ under $y_i$ and $z_i$, we get $P_s$ in (4) (note: for a given $y_i$, $rc_i = 1,...,y_i$ with equal probability; similarly, for a given $z_i$, $rt_i = 1,...,z_i$ with equal probability). In other words, if (C4) is valid, the transmission-state distribution $P_s$ remains the same as that of exponential distributions under mixtures of gamma distributions for countdown and transmission times.

**Proof**: The proof is similar to that of Theorem C1. Therefore, we just sketch the proof here rather than repeating the mechanic. We could imagine that there are multiple layers of states in the state-transition diagram, with each layer consisting of states corresponding to a specific combination of $y_i$'s and $z_i$'s. The state moves within the same layer when the transmission state $s$ does not change during transition. The transition rates within each layer are the same as before for both the forward and reverse processes. When $s$ changes due to a change of $s_i$ of a link $i$, then the state may move to another layer depending on the new $z_i$' generated (if $s_i$ jumps from 0 to 1), or the new $y_i'$ generated (if $s_i$ jumps from 1 to 0). The transition rates from the old state $x$ to the new state $x'$ are $p_Z(z_i')/d$ and $p_Y(y_i')/e$, respectively. The corresponding transition rates from $x'$ to $x$ for the reverse processes are $p_Y(y_i)/e$ and $p_Z(z_i)/d$. We then go through the three cases as in the proof of Theorem C2. □

*Approximating arbitrary distributions by taking limits on mixtures of gamma distributions*

Now suppose the distributions $P_Y(y)$ and $P_Z(z)$ are concentrated around large values of $y$ and $z$. For each realization $y$, $E[T_{cd}|y] = yd$, $Var(T_{cd}|y) = yd^2$. If we keep $E[T_{cd}|y] = yd$ constant while letting $y \to \infty$, then $Var(T_{cd}|y) \to 0$. Thus, $T_{cd}$ given $y$ becomes a deterministic value. By the same token, for large $z$, $T_{tr}$ given $z$ becomes a deterministic value under similar limit. Thus, each $P_Y(y)$ or $P_Z(z)$ is mapped to each $P_{T_{cd}}(t_{tr})$ or $P_{T_{tr}}(t_{tr})$, respectively. This is an approximation for arbitrary discrete distributions of $T_{cd}$ and $T_{tr}$. For an approximation for arbitrary continuous distributions $f(t_{tr})$ and of $g(t_{tr})$, we could make the number of possible values for $Y$ and $Z$ very large. In either case, $P_s$ given by (4) remains valid under Theorem C2.